\documentstyle[aps,prl,preprint,floats,epsfig]{revtex}

\begin{document}

\preprint{\tighten\vbox{\hbox{\hfil CLNS 00/1670}
                        \hbox{\hfil CLEO 00-7}
}}

\title{\bf Measurement of the Relative Branching Fraction of $\Upsilon(4S)$ 
to Charged and Neutral $B$-Meson Pairs}

\author{CLEO Collaboration}
\date{\today}

\maketitle
\tighten

\begin{abstract} 

We analyze $9.7 \times 10^6 B\overline{B}$
pairs recorded with the CLEO detector to determine the production ratio 
of charged to neutral $B$-meson pairs produced at the $\Upsilon$(4S) resonance.
We measure the rates 
for $B^0 \to J/\psi K^{(*)0}$ and $B^+ \to J/\psi K^{(*)+}$ decays and use
the world-average $B$-meson lifetime ratio to extract the relative widths
$\frac{f_{+-}}{f_{00}} =
\frac{\Gamma(\Upsilon(4\rm{S}) \to B^{+}B^{-})}
     {\Gamma(\Upsilon(4\rm{S}) \to B^{0}\overline{B}^{0})}
= 1.04  \pm 0.07(stat) \pm 0.04(syst)$.
With the assumption that $f_{+-}+f_{00}=1$, we obtain
$f_{00} = 0.49 \pm 0.02(stat) \pm 0.01(syst)$ and
$f_{+-} = 0.51 \pm 0.02(stat) \pm 0.01(syst)$.
This production ratio and its uncertainty apply to all
exclusive $B$-meson branching fractions measured at the $\Upsilon$(4S) 
resonance. 
\end{abstract}

\newpage

{
\renewcommand{\thefootnote}{\fnsymbol{footnote}}

\begin{center}
J.~P.~Alexander,$^{1}$ R.~Baker,$^{1}$ C.~Bebek,$^{1}$
B.~E.~Berger,$^{1}$ K.~Berkelman,$^{1}$ F.~Blanc,$^{1}$
V.~Boisvert,$^{1}$ D.~G.~Cassel,$^{1}$ M.~Dickson,$^{1}$
P.~S.~Drell,$^{1}$
% ...
\end{center}
 
\small
\begin{center}
$^{1}${Cornell University, Ithaca, New York 14853}\\
% ...
\end{center}
 
\setcounter{footnote}{0}
}
\newpage

Measurements of exclusive $B$-decay branching fractions from $e^+e^-$ collider
operation at the $\Upsilon$(4S) resonance assume equal production 
rates of charged and neutral $B$-meson pairs~\cite{PDG98}.
In the literature, the uncertainty in a specific branching fraction due to a 
lack of knowledge of the production ratio is often ignored. 

Any physics based upon comparisons of absolute decay
rates of charged and neutral $B$ mesons will profit from a more precise 
knowledge of the $B$-production ratio, 
$f_{+-}/f_{00}\equiv 
\Gamma(\Upsilon(4S) \to B^+B^-)/
\Gamma(\Upsilon(4S) \to B^0\overline{B}^0)$.
For example, a comparison of the
branching fractions of two-body hadronic decays can be used to obtain 
information on
the relative contributions from external and internal
spectator decays~\cite{HadronicBDecay}. For all exclusive decay modes
studied, the $B^+$ branching fraction was found to be larger than the
corresponding $B^0$ branching fraction, indicating constructive
interference between the external and internal spectator amplitudes.
This is in contrast to the destructive interference observed
in hadronic charm decay. The magnitude of the constructively
interfering fraction depends on the value of $f_{+-}/f_{00}$.
Another application of the $f_{+-}/f_{00}$ ratio arises in 
the use of ratios of charmless hadronic 
$B$-decay rates~\cite{Buras_Neubert_Soni} to
set bounds on the angle $\gamma$, the phase of the CKM matrix element 
$V_{ub}$~\cite{PDG98,CKM}. 
The uncertainty on $f_{+-}/f_{00}$ contributes to the systematic uncertainty 
of the $\gamma$ bound.

A better measurement of $f_{+-}/f_{00}$ would also allow a 
more meaningful comparison with theoretical predictions of the 
relative $B^+B^-$ and $B^0\overline{B}^0$ production rates at the
$\Upsilon$(4S) resonance. 
If there are no other important
differences between the two $\Upsilon$(4S) decays, such as $B^+-B^0$
mass splitting or isospin-violating form factors in the decay amplitude,
Coulomb corrections
to $B^+B^-$ production near threshold are not negligible, giving rise to
$\frac{\Gamma(\Upsilon(4S) \to B^+B^-)}
      {\Gamma(\Upsilon(4S) \to B^0\overline{B}^0)} \simeq 1.18$~\cite{Marciano}.
Other authors~\cite{Byers_Lepage} argue that the $B$-meson substructure cannot
be ignored and strongly
reduces the Coulomb effect in the $B$-production ratio 
to 1.05$-$1.07, depending on the $B$ masses and momenta.

Existing measurements of the admixture ratio of charged to
neutral $B$ mesons produced at the $\Upsilon(4S)$ resonance 
have an uncertainty of 
$\sim$15\%. One measurement~\cite{cleo96_1455}
used the branching-fraction ratio of ${\cal B}(B^+ \to J/\psi K^{(*)+})$ to 
${\cal B}(B^0 \to J/\psi K^{(*)0})$~\cite{Kstar_ref} to yield
$\frac{f_{+-}}{f_{00}} \times \frac{\tau_{B^+}}{\tau_{B^0}} 
= 1.15 \pm 0.17 \pm 0.06$, where the first uncertainty is statistical, the
second is systematic, and $\tau_B$ denotes the $B$ lifetime.
Another measurement~\cite{sl_cleo95} 
used a ratio of $B \to D^* l \nu$ decays to extract
$ \frac{f_{+-}}{f_{00}} \times \frac{\tau_{B^+}}{\tau_{B^0}} 
= 1.14 \pm 0.14 \pm 0.13$.

In the present analysis, we study the decays 
$B \to J/\psi K^{(*)}$, which are isospin conserving
transitions, since the $J/\psi$ daughter is an
iso-singlet and the $B$ and $K^{(*)}$ mesons are both iso-doublets. The
decays $B^+ \to J/\psi K^{(*)+}$ and $B^0 \to J/\psi K^{(*)0}$
must therefore have equal partial widths and we can extract
$R \equiv \frac{f_{+-}}{f_{00}} \times \frac{\tau_{B^+}}{\tau_{B^0}}
= \frac{{\cal N}(B^+ \to J/\psi K^{(*)+})}
          {{\cal N}(B^0 \to J/\psi K^{(*)0})}$, where ${\cal N}$ is the
efficiency-corrected signal yield.
Using the ratio of two similar decay rates to extract $R$, we exploit
the cancellation of common experimental uncertainties.
Throughout this Letter, reference to charge conjugate 
states is implicit.

The data analyzed in this study 
were recorded at the Cornell Electron Storage Ring (CESR) with two 
configurations of the CLEO detector, CLEO~II and CLEO~II.V. 
The data consist of an integrated luminosity of $9.2~$fb$^{-1}$ of $e^+e^-$ 
annihilations recorded at the
$\Upsilon(4S)$ resonance and of $4.6~$fb$^{-1}$ taken in the 
continuum, 60~MeV below the $\Upsilon(4S)$ energy.
The results in this Letter are based upon $9.7~\times~10^6$
$B\overline{B}$ candidates and supersede those of Ref.~\cite{cleo96_1455}.

The components of the CLEO detector most relevant to this analysis are the
charged-particle tracking system, the 7800-crystal CsI electromagnetic 
calorimeter, and the muon chambers. 
The first third of the data were collected with the CLEO~II 
detector~\cite{CLEOII}, which measured the momenta of charged particles 
in a tracking system consisting of an inner 6-layer straw-tube chamber, 
a 10-layer 
precision drift chamber, and a 51-layer main drift chamber, all operating
inside a 1.5~T solenoidal magnet. The main drift chamber also provided
a measurement of the specific ionization loss ($dE/dx$) used in particle
identification. 
Two thirds of the data were taken with the CLEO~II.V 
configuration, for which
the innermost straw-tube chamber was replaced with
a 3-layer silicon vertex detector~\cite{CLEOII.V}, and the argon-ethane gas
of the main drift chamber was replaced with a helium-propane mixture. 
The muon identification system in both the CLEO~II and CLEO~II.V 
configurations consisted of proportional counters placed at 
various depths in the return yoke of the magnet.

Since the backgrounds for $B \to J/\psi K^{(*)}$ decays are very low,
track and photon quality requirements have been designed to maximize 
signal yield.
We reconstruct $B \to J/\psi K^{(*)}$ candidates 
in the data samples taken at the $\Upsilon(4S)$ energy. 

Candidate $J/\psi$ mesons are reconstructed in their leptonic decay modes,
requiring $J/\psi$ lepton daughter tracks to have momenta larger than 
800~MeV/$c$.

For $J/\psi$ reconstruction in the muon channel, one of the muon 
candidates was required to penetrate the steel absorber to a depth greater 
than three nuclear interaction lengths. For the opposite sign daughter 
candidate, no muon detection requirement was imposed.

Electron candidates
were identified based on the ratio of the track momentum to the associated 
shower energy in the CsI calorimeter and specific ionization loss in the
drift chamber.
Bremsstrahlung produces a radiative tail in the $e^+e^-$ invariant mass 
distribution below the $J/\psi$ pole. We recovered some of the resultant 
efficiency loss by detecting the radiated photon.
We selected photon candidates ($E_\gamma > $~10 MeV)
with the smallest angle to the $e^{\pm}$ track, provided this angle
did not exceed 5$^{\circ}$. The $J/\psi \to e^+e^-$ efficiency was increased by
$\sim$20\%, without adding background.

\begin{figure*}[t]
\begin{center}
        \epsfxsize 4.0in \epsffile{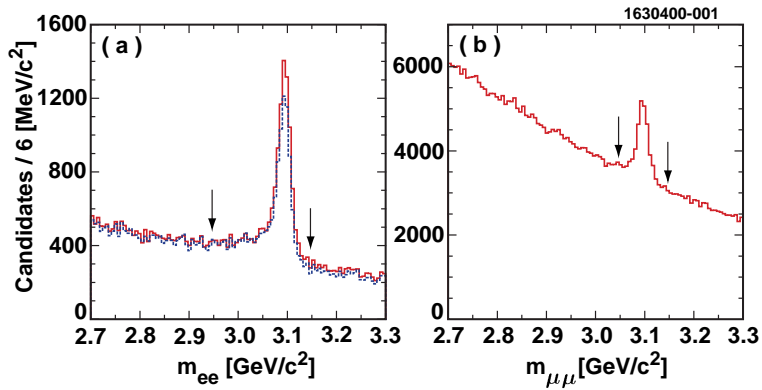}
\end{center}
\caption{Invariant mass spectrum of the (a) $J/\psi \to e^+e^-$ and (b) 
$J/\psi \to \mu^+\mu^-$ candidates. The dashed line in (a) shows the mass 
spectrum before the addition of bremsstrahlung photons. The arrows
delimit the $J/\psi$ candidate region.}
\label{Psi_Incl}
\end{figure*}

We reconstructed $15\,900 \pm 700$ 
inclusive $J/\psi \to l^+l^-$ candidates
(Fig.~\ref{Psi_Incl}), about equally shared in the two dilepton 
reconstruction modes. The resolution in the $J/\psi$ 
invariant mass was $\sim$13~MeV.

We required the dimuon invariant mass to be within 50~MeV of the
world-average $J/\psi$ mass~\cite{PDG98}, 
corresponding to a $\sim$3.5 standard deviation ($\sigma$) 
selection. For the dielectron invariant mass we required 
$-150$~MeV $< (m_{ee} - m_{J/\psi}) <$ 50~MeV to 
allow for the radiative tail. The $J/\psi$ energy resolution was improved by a
factor $\sim$4 after performing a kinematic fit of the 
dilepton invariant mass to the $J/\psi$ mass. We required $J/\psi$ 
candidates to have momenta below 2~GeV/$c$, which is near 
the kinematic limit for $J/\psi$ mesons
originating from a $B$ meson nearly at rest.

The $K^0_S \to \pi^+\pi^-$ candidates were selected from pairs of tracks forming well-measured
displaced vertices. The resolution in $\pi^+\pi^-$ invariant mass is 
approximately 2.5~MeV. Due to very low background in $B \to J/\psi K^0_S$ 
candidates, we only require that
neutral kaon candidates have a normalized mass within $10\sigma$ 
(because the $K^0_S$ mass distribution has non-negligible non-Gaussian tails)
and a normalized flight distance greater than zero.

Charged kaon and pion candidates are required to have a measured $dE/dx$ 
within $3\sigma$ of the energy loss expected for the given particle type. 
Neutral pions
are reconstructed from photon pairs detected within the barrel region
of the CsI calorimeter, 
$\mid \cos \theta_\gamma \mid < 0.71$, where $\theta_\gamma$ is the polar 
angle of the candidate photon with respect to the $e^+e^-$ 
beam axis. The photons 
must have a minimum energy of 30~MeV and their normalized invariant mass is 
required to be within $2.5 \sigma$ of the
$\pi^0$ mass. This diphoton invariant mass 
is then kinematically constrained to the $\pi^0$ mass.
Charged and neutral pions and kaons are used to reconstruct the four $K^*$ 
decay modes. 
Candidate $K^*$ mesons are required to have a $K\pi$ invariant mass within
75~MeV of the world-average $K^*$ mass~\cite{Kstar_ref}. 

\begin{figure*}[htp]
\begin{center}
        \epsfxsize 5.0in \epsffile{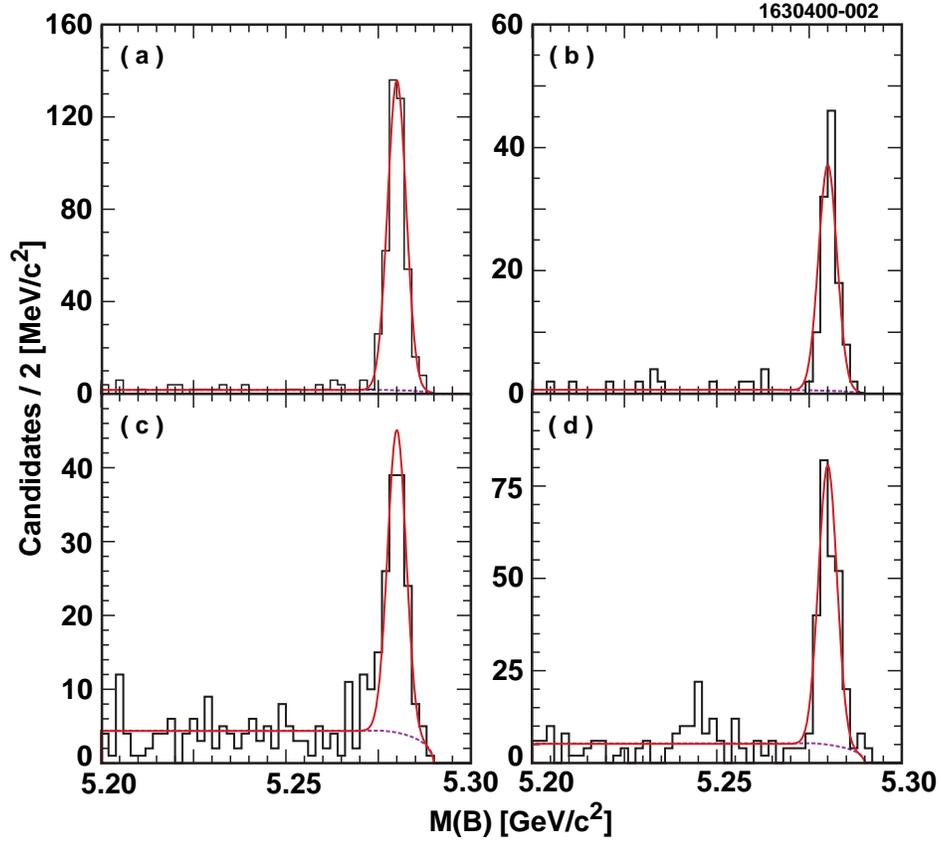}
\end{center}
\caption{Beam-constrained mass projections (histograms) 
for candidates in the $\Delta E$ 
signal region are shown for the entire data sample summed over
both $J/\psi \to l^+l^-$ modes. The fits to the data are shown 
with the solid curves 
while the background fits are given with the dashed curves.
Shown are distributions for (a) $B^+ \to J/\psi K^+$, 
(b) $B^0 \to J/\psi K^0_S$, (c) $B^+ \to J/\psi K^{*+}$,
and (d) $B^0 \to J/\psi K^{*0}$ candidates.}
\label{dE_mbc}
\end{figure*}

We fully reconstruct $B$-meson candidates by employing the kinematics of a 
$B\overline{B}$ pair produced almost at rest. We use the energy difference 
$\Delta E \equiv E(J/\psi) + E(K^{(*)}) - E_{\rm{beam}}$ as well as the
beam-constrained mass $M(B) \equiv \sqrt{E^2_{\rm{beam}} - p^2(B)}$ 
as selection observables. The resolution in $\Delta E$ is 15~MeV for 
$J/\psi K^*$ with a $\pi^0$ candidate 
in the final state and 9$-$11 MeV for the other modes.
We find the resolution in $M(B)$ to be $\sim$2.5~MeV,
which is dominated by the beam energy spread. 
We select signal candidates by requiring $5.2$~GeV$ < M(B) <$ 5.3~GeV 
and $\mid \Delta E \mid < 3\sigma_{\Delta E}$.
The beam-constrained mass distributions for events within the 
$\Delta E$ signal region are shown in Fig.~\ref{dE_mbc}. 

We extract the signal yield in each mode by performing a binned 
maximum-likelihood fit to the $M(B)$ projection, where the signal
is given by a single Gaussian distribution with fixed mean of 
5.28~GeV and fixed width of 2.5~MeV. The background is fit to a
first-order polynomial joined with an elliptic function to fit the threshold 
nature of the beam-constrained mass distribution.
The $M(B)$ distributions in the $\Delta E$ sideband regions exhibit a 
slope consistent with zero.
These sideband regions are at least $4~\sigma_{\Delta E}$
and less than one pion mass away from the $\Delta E$ signal region.
We fix the slope of the background shape to zero and allow the level of the 
combinatoric background to be determined from the fit to the $M(B)$ projection
of the $\Delta E$ signal region.

We must account for the individual final
states being reconstructed in a different channel (cross-feed), since for such 
candidates both the total
energy and the beam-constrained mass lie near the signal region.
We evaluate the reconstruction efficiency, as well as the amount of 
cross-feed from a given channel {\it i} to another channel {\it j}, 
using a sample of simulated $B \to J/\psi K^{(*)}_{\it i}$ events
to generate a $6 \times 6$ efficiency matrix for the
$J/\psi \to e^+e^-$ and $J/\psi \to \mu^+\mu^-$ cases, as well as for 
CLEO~II and CLEO~II.V, separately. 
The CLEO detector simulation is based upon GEANT~\cite{GEANT}. 
Simulated events are processed in a manner similar to that for the data. 
There is negligible cross-feed between 
the $J/\psi K$ and the $J/\psi K^*$ modes. The cross-feed 
into $J/\psi K^*$ modes with a charged-pion $K^*$ daughter is near 5\%, 
whereas cross-feed into $J/\psi K^*$ modes with a neutral-pion $K^*$ daughter
ranges between $8-30$\% of the raw yield.
Efficiencies and
cross-feed-corrected yields are listed in Table~\ref{eff_table}. 
As a cross check, we also quote
the branching fractions computed for the analyzed $B \to J/\psi K^{(*)}$ 
decays. 

\begin{table*}[htbp]
\tiny
\begin{center}
\begin{tabular}{lccccc} 
               & \multicolumn{2}{c}{efficiency [\%]} 
               & \multicolumn{2}{c}{cross-feed corrected yield} 
               & $\cal{B}$\\
               & $J/\psi \to e^+e^-$ & $J/\psi \to \mu^+\mu^-$ 
               & $J/\psi \to e^+e^-$ & $J/\psi \to \mu^+\mu^-$
               & $[\times 10^{-3}]$ \\
\hline
\multicolumn{6}{c}{CLEO~II}\\
$J/\psi K^+$ 
               & $46.0 \pm 0.7$ & $56.2 \pm 0.7$
               & $87.6 \pm 9.4 \pm 1.3$ & $121.9 \pm 11.3 \pm 1.5$
               & $1.02 \pm 0.07$ \\
$J/\psi K^0_S$ 
               & $43.3 \pm 0.6$ & $51.9 \pm 0.7$
               & $32.9 \pm 5.7 \pm 0.5$ 
               & $\phantom{0}24.0 \pm \phantom{0}5.1 \pm 0.3$
               & $0.83 \pm 0.12$ \\
$J/\psi K^{*+}(K^0_S\pi^+)$ 
               & $28.0 \pm 0.6$ & $34.1 \pm 0.7$
               & $13.1 \pm 3.8 \pm 0.3$ 
               & $\phantom{0}15.9 \pm \phantom{0}4.6 \pm 0.3$
               & $1.02 \pm 0.21$ \\
$J/\psi K^{*+}(K^+\pi^0)$ 
               & $16.5 \pm 0.4$ & $20.7 \pm 0.4$
               & $17.1 \pm 4.7 \pm 0.7$ 
               & $\phantom{0}23.6 \pm \phantom{0}6.2 \pm 0.5$
               & $1.63 \pm 0.31$ \\
$J/\psi K^{*0}(K^+\pi^-)$
               & $33.5 \pm 0.5$ & $40.6 \pm 0.5$
               & $53.1 \pm 7.5 \pm 0.8$ 
               & $\phantom{0}56.8 \pm \phantom{0}8.0 \pm 0.7$
               & $1.11 \pm 0.11$ \\
$J/\psi K^{*0}(K^0_S\pi^0)$ 
               & $16.4 \pm 0.4$ & $18.7 \pm 0.4$
               & $3.4 \pm 2.3 \pm 0.2$ 
               & $\phantom{0}\phantom{0}5.1 \pm \phantom{0}3.0 \pm 0.2$
               & $1.02 \pm 0.46$ \\
\hline
\multicolumn{6}{c}{CLEO~II.V}\\
$J/\psi K^+$
               & $43.6 \pm 0.7$ & $58.8 \pm 0.8$
               & $172.5 \pm 13.1 \pm 2.8$ & $210.2 \pm 15.0 \pm 2.9$
               & $0.98 \pm 0.05$ \\
$J/\psi K^0_S$
               & $42.5 \pm 0.5$ & $57.9 \pm 0.6$
               & $\phantom{0}42.5 \pm \phantom{0}6.5 \pm 0.6$ 
               & $\phantom{0}78.9 \pm \phantom{0}9.1 \pm 0.9$
               & $0.90 \pm 0.08$ \\
$J/\psi K^{*+}(K^0_S\pi^+)$
               & $28.6 \pm 0.6$ & $36.4 \pm 0.7$
               & $\phantom{0}17.0 \pm \phantom{0}4.6 \pm 0.5$ 
               & $\phantom{0}47.6 \pm \phantom{0}7.3 \pm 0.9$
               & $1.00 \pm 0.14$ \\
$J/\psi K^{*+}(K^+\pi^0)$ 
               & $14.8 \pm 0.4$ & $21.6 \pm 0.4$
               & $\phantom{0}35.9 \pm \phantom{0}6.8 \pm 1.0$ 
               & $\phantom{0}42.0 \pm \phantom{0}8.2 \pm 1.0$
               & $1.69 \pm 0.23$ \\
$J/\psi K^{*0}(K^+\pi^-)$ 
               & $31.1 \pm 0.7$ & $41.1 \pm 0.8$
               & $\phantom{0}92.5 \pm \phantom{0}9.8 \pm 2.1$ 
               & $105.8 \pm 11.1 \pm 2.1$
               & $1.08 \pm 0.08$ \\
$J/\psi K^{*0}(K^0_S\pi^0)$ 
               & $15.5 \pm 0.5$ & $18.2 \pm 0.5$
               & $\phantom{0}11.9 \pm \phantom{0}4.0 \pm 0.5$ 
               & $\phantom{0}\phantom{0}8.6 \pm \phantom{0}4.0 \pm 0.4$
               & $1.37 \pm 0.38$ \\
\end{tabular}
\end{center}
\caption{Summary of reconstruction efficiencies (daughter branching 
fractions not included), cross-feed corrected signal yields (first error
is statistical, second error is systematic), and branching fractions ${\cal B}$
computed from these yields and efficiencies 
(errors are statistical only), assuming equal 
production of $B^+B^-$ and $B^0\overline{B}^0$ pairs. The computed branching
fractions agree with the world-average values~\protect\cite{PDG98}.
Results for the data accumulated with the CLEO~II and CLEO~II.V configurations
are given separately.}
\label{eff_table}
\end{table*}

We extract our result from the cross-feed and reconstruction-efficiency 
corrected yields using the world-average values for the respective daughter 
decay branching fractions~\cite{PDG98}. 
We obtain four independent measurements of $R$ listed in 
Table~\ref{tabResults}.

\begin{table}[htp]
\begin{center}
\begin{tabular}{lcc}
~~~~~~~~~Configuration \\
\unitlength1pt
\begin{picture}(0,0)
\put(0,18){\line(3,-1){90}} 
\end{picture}
\phantom{0} & CLEO~II   & CLEO~II.V \\
Signal Mode &   & \\
\hline
$B \to J/\psi K$   & $1.229 \pm 0.191$ & $1.088 \pm 0.116$ \\
$B \to J/\psi K^*$ & $1.098 \pm 0.190$ & $1.095 \pm 0.137$ \\
\end{tabular}
\end{center}
\caption{Results for 
$R = \frac{f_{+-}}{f_{00}} \times \frac{\tau_{B^+}}{\tau_{B^0}}$
for the different CLEO configurations and the $J/\psi K$ and $J/\psi K^*$
modes. The uncertainties are statistical.}
\label{tabResults}
\end{table}

We evaluate the uncertainties in the reconstruction efficiency due to 
track finding, track fitting, charged hadron identification, 
$K^0_S$ finding, and 
$\pi^0$ finding. Since we use the ratio of two decay rates 
that each involve $J/\psi \to l^+l^-$ candidates, 
uncertainties in lepton identification 
are negligible.
We estimate the full systematic bias due to daughter reconstruction
efficiency uncertainties by taking into account correlations between the 
different final states in the numerator and denominator, 
resulting in some cancellation. 
Propagating these uncertainties through the weighted average of the results
in Table~\ref{tabResults},
we arrive at a systematic uncertainty on $R$ due to the understanding of 
reconstruction efficiencies of $^{+1.0\%}_{-1.5\%}$.
The polarization of the decay $B \to J/\psi K^*$ is modeled in our simulation with
a longitudinal polarization fraction of $\Gamma_L/\Gamma$ = 0.52 in accordance 
with Ref.~\cite{cleo96_1455}. We estimate the impact of the value used 
for $\Gamma_L/\Gamma$ on the $B \to J/\psi K^*$ efficiencies by generating signal
events with the nominal polarization varied by $\pm 1\sigma=\pm0.08$. 
The central value for $R$ changes by less than 0.8\% due to this variation.
We vary the $B$ candidate signal width by $\pm$0.2~MeV and
estimate the systematic uncertainty from this source to be less than 0.5\%.
We extract the signal yield using a background function that allows for a slope
in the non-signal region of the beam-constrained mass distribution and
assign a systematic bias of 3\% on the central value for $R$ 
from our assumption of flat background.
We attribute a 1.1\% uncertainty to limited statistics of the simulated 
event samples used to extract the efficiency matrices.
Adding all contributions in quadrature we obtain a total
systematic uncertainty of $^{+3.5\%}_{-3.7\%}$ on $R$.

We weight the results of Table~\ref{tabResults} with their statistical 
uncertainty 
and, combined with the estimated systematic uncertainty, we extract

\begin{center}
$R \equiv \frac{f_{+-}}{f_{00}} \times \frac{\tau_{B^+}}{\tau_{B^0}} 
= 1.11 \pm 0.07 \pm 0.04, $ \\
\end{center}

where the first uncertainty is statistical and the second is systematic.

Using the world-average lifetime ratio of charged and neutral $B$ mesons, 
$1.066 \pm 0.024$~\cite{B_lifetime_group}, we obtain a measurement of the
production ratio

\begin{center}
$ \frac{f_{+-}}{f_{00}}  = \frac{\Gamma(\Upsilon(4\rm{S}) \to B^{+}B^{-})}
  {\Gamma(\Upsilon(4\rm{S}) \to B^{0}\overline{B}^{0})}
                        = 1.04 \pm 0.07 \pm 0.04, $ \\
\end{center}

and, assuming $f_{+-} + f_{00} = 1$, we also extract
$f_{00}  = 0.49 \pm 0.02 \pm 0.01$ and
$f_{+-} = 0.51 \pm 0.02 \pm 0.01$.

We have measured the ratio of charged to neutral production
of $B$ mesons at the $\Upsilon(4S)$ resonance~\cite{refPeak}
to be $1.04 \pm 0.07 \pm 0.04$, which is consistent with unity within
an error of 8\%. This is the most precise measurement
of $f_{+-}/f_{00}$. Our result is also consistent with theoretical
predictions of greater charged than neutral $B$-meson production 
in $\Upsilon(4S)$ decays near threshold.
We emphasize that the ratio $f_{+-}/f_{00}$ 
and its uncertainty must be taken into account when performing
measurements that compare charged and neutral $B$ decays at the 
$\Upsilon(4S)$ resonance.

We gratefully acknowledge the effort of the CESR staff in providing us with
excellent luminosity and running conditions.
This work was supported by 
the National Science Foundation,
the U.S. Department of Energy,
the Research Corporation,
the Natural Sciences and Engineering Research Council of Canada, 
the A.P. Sloan Foundation, 
the Swiss National Science Foundation, 
and the Alexander von Humboldt Stiftung.


\begin{thebibliography}{99}
\bibitem{refPeak}
We took most of our data near the central maximum of the $\Upsilon(4S)$ 
energy resonance; we note, however, that $f_{+-}/f_{00}$ 
could vary as a function of energy in the $\Upsilon(4S)$ resonance region.
\end{thebibliography}
\end{document}